\documentclass[aps,prl,preprint,amsmath,amssymb,showpacs,preprintnumbers,superscriptaddress]{revtex4-1}
\usepackage{graphicx}
\usepackage{dcolumn}
\usepackage{bm}

\usepackage{hyperref}

\begin{document}


\title{Rod-shaped Nuclei at Extreme Spin and Isospin}

\author{P. W. Zhao 
}
\affiliation{Yukawa Institute for Theoretical Physics, Kyoto University, Kyoto 606-8502, Japan}
\affiliation{Physics Division, Argonne National Laboratory, Argonne, Illinois 60439, USA}
\affiliation{State Key Laboratory of Nuclear Physics and Technology, School of Physics, 
Peking University, Beijing 100871, China}

\author{N. Itagaki 
}
\affiliation{Yukawa Institute for Theoretical Physics, Kyoto University, Kyoto 606-8502, Japan}

\author{J. Meng 
}
\affiliation{State Key Laboratory of Nuclear Physics and Technology, School of Physics, 
Peking University, Beijing 100871, China}
\affiliation{School of Physics and Nuclear Energy Engineering, Beihang University, Beijing 100191, China}
\affiliation{Department of Physics, University of Stellenbosch, Stellenbosch 7602, South Africa}


\begin{abstract}
The anomalous rod shape in carbon isotopes has been investigated in the framework of the cranking covariant density functional theory, and two mechanisms to stabilize such a novel shape with respect to the bending motion, extreme spin, and isospin, are simultaneously discussed for the first time in a self-consistent and microscopic way. 
By adding valence neutrons and rotating the system, we have found the mechanism stabilizing the rod shape; i.e., the $\sigma$ orbitals (parallel to the symmetry axis) of the valence neutrons, important for the rod shape, are lowered by the rotation due to the Coriolis term. The spin and isospin effects enhance the stability of the rod-shaped configuration. This provides a strong hint that a rod shape could be realized in nuclei towards extreme spin and isospin.
\end{abstract}


\maketitle



Strong nuclear deformations provide us an excellent framework to
investigate the fundamental properties of quantum many-body systems. 
Experiments have provided evidence in heavy nuclei for the existence of the so-called super-~\cite{Nyako1984Phys.Rev.Lett.507,Twin1986Phys.Rev.Lett.811} and hyperdeformation~\cite{Galindo-Uribarri1993Phys.Rev.Lett.231,LaFosse1995Phys.Rev.Lett.5186,Krasznahorkay1998Phys.Rev.Lett.2073}, i.e., strong deformation with width-to-length ratios of 1:2 or 1:3.
For light nuclei, there have been indications that even more exotic states above 1:3 might exist in light $N=Z$ nuclei due to the $\alpha$ cluster structure. However, there is still no firm evidence so far, despite intensive experimental searches.

The realization of an anomalously deformed rod shape in light nuclei has been a long-standing objective of nuclear structure physics. Because of the antisymmetrization effects and the weak-coupling nature, it has been known to be difficult to stabilize the rod-shaped configuration in nuclear systems. The linear-chain structure of three-$\alpha$ clusters was suggested about 60 years ago~\cite{Morinaga1956Phys.Rev.254} and was used to explain the structure of the Hoyle state (the second $0^+$ state at $E_x = 7.65$ MeV in $^{12}$C), which plays a crucial role in the synthesis of $^{12}$C from three $^4$He nuclei in stars~\cite{Hoyle1954Astrophys.J.Suppl.Ser.121}. However, this state was later found to be a gaslike state with strong mixing of the linear-chain configuration and various other three-$\alpha$ configurations~\cite{Fujiwara1980Prog.ofTheor.Phys.Suppl.29} and recently reinterpreted as an $\alpha$-condensate-like state~\cite{Tohsaki2001Phys.Rev.Lett.192501,Suhara2014Phys.Rev.Lett.62501}. Therefore, various theoretical and experimental studies of linear-chain states have been carried out in other $N=Z$ nuclei~\cite{Freer2007Rep.Prog.Phys.2149} such as $^{16}$O~\cite{Chevallier1967Phys.Rev.827,Suzuki1972Prog.Theor.Phys.1517,Flocard1984Prog.Theor.Phys.1000,Bender2003Nucl.Phys.A390,Ichikawa2011Phys.Rev.Lett.112501,Yao2014Phys.Rev.C54307}, $^{24}$Mg~\cite{Iwata2014,Wuosmaa1992Phys.Rev.Lett.1295}, etc.; further investigations are needed to confirm, however.

To stabilize the linear-chain configuration with respect to the bending motion, some extra mechanisms need to be introduced. One of the candidates is the increase of isospin by adding valence neutrons. 
Even if the linear-chain configurations are difficult to stabilize in $N=Z$ nuclei, higher stability is possible in the neutron-rich side. In particular,
if the neutrons occupy the so-called $\sigma$ orbital (orbital parallel to the symmetry axis), an elongated shape for the core would be favored to lower the energy of the valence neutrons~\cite{Itagaki2001Phys.Rev.C14301,Itagaki2004Phys.Rev.Lett.142501}. 
This is because originally $\sigma$ orbitals are higher nodal orbitals, but their energies are lowered by the prolate deformation. Eventually, prolate deformation is induced when the neutrons occupy the $\sigma$ orbitals. The effects of the valence neutrons on cluster structure have been extensively investigated from both experimental~\cite{Freer2006Phys.Rev.Lett.42501,Navin2000Phys.Rev.Lett.266} and theoretical sides~\cite{Itagaki2000Phys.Rev.C44306,Ito2008Phys.Rev.Lett.182502,Maruhn2010Nucl.Phys.A1,Baba2014Phys.Rev.C64319}.
Another possible mechanism is the increase of the angular momentum by rotating the nucleus rapidly, because the linear-chain configuration with a large moment of inertia should be favored with a large angular momentum. In this case, the competition between the nuclear attractive and centrifugal forces~\cite{Wilkinson1986Nucl.Phys.A296} would be very important for the stabilization of the linear-chain state. 

Until now, most of the theoretical analyses of the linear-chain structure have been performed by using the conventional cluster model with effective interactions determined from the binding energies and scattering phase shifts of the clusters~\cite{VonOertzen2006Phys.Rep.43}. Therefore, it is highly desirable to have investigations based on different approaches, such as density functional theories (DFTs). Since the DFTs do not {\it a priori} assume the existence of $\alpha$ clusters, it would provide more confidence in the presence of exotic cluster structure as a result of calculations. Such calculations are not easy and, so far, have only rare examples including the linear-chain configurations of $^{16,20}$C~\cite{Maruhn2010Nucl.Phys.A1}, $^{16}$O~\cite{Flocard1984Prog.Theor.Phys.1000,Bender2003Nucl.Phys.A390,Ichikawa2011Phys.Rev.Lett.112501,Yao2014Phys.Rev.C54307}, and other light $N=Z$ nuclei~\cite{Ebran2014Phys.Rev.C54329}. 
Even now, the knowledge on the stabilization of the linear-chain state is insufficient. 
To clarify the nature of linear-chain states, it is important to explore the two mechanisms of large isospin and high spin in their stabilization.

The cranking model~\cite{Inglis1956Phys.Rev.1786} is a reliable method for the description of states with good angular momentum. 
It is a first-order approximation for a variation after projection onto good angular momentum~\cite{Beck1970Z.Phys.26}, and has been extended to provide a very successful self-consistent description of rotational nuclei all over the periodic table.
The covariant density functional theory (CDFT) exploits basic properties of QCD at low energies, in particular, symmetries and the separation of scales~\cite{Lalazissis2004_LNP641x}. 
CDFT consistently treats the spin degrees of freedom, includes the complicated interplay between the large Lorentz scalar and vector self-energies induced on the QCD level~\cite{Cohen1992Phys.Rev.C1881}, and naturally provides the nuclear currents induced by the spatial parts of the vector self-energies, which play an essential role in rotating nuclei. 
The cranking CDFT~\cite{Peng2008Phys.Rev.C24313,Zhao2011Phys.Lett.B181,Meng2013FrontiersofPhysics55} has provided an excellent description of ground states and the rotational excited states all over the periodic table with a high predictive power~\cite{Zhao2011Phys.Rev.Lett.122501,Zhao2012Phys.Rev.C54310}. It has been shown recently that relativistic models are especially suited for the self-consistent microscopic description of cluster phenomena in nuclei~\cite{Ebran2012Nature341}.

In this Letter, both mechanisms, adding neutrons and rotating the system, are taken into account in a microscopic and self-consistent way for the first time for the stability of the linear-chain state with respect to the bending motion. The cranking covariant DFT~\cite{Peng2008Phys.Rev.C24313,Zhao2011Phys.Lett.B181,Meng2013FrontiersofPhysics55} will be used to investigate the stability of the anomalously deformed rod shape in C isotopes toward the extreme isospin and spin.

\begin{figure}[!htbp]
\centering
\includegraphics[width=6.5cm]{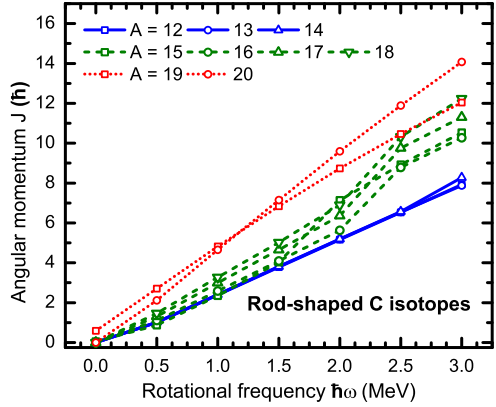}
\caption{(color online) Angular momenta as functions of the rotational frequency for C isotopes from $A=12$ to $A=20$.}
\label{fig1}
\end{figure}

The covariant DFT starts from a Lagrangian, and the corresponding Kohn-Sham equations have the form of a Dirac equation with effective fields $S(\bm{r})$ and $V^\mu(\bm{r})$ derived from this Lagrangian. In the cranking model, these fields are deformed, and the calculations are carried out in the intrinsic frame rotating with a constant angular velocity vector $\bm{\omega}$, which, in this work, points in a direction perpendicular to the symmetry axis $z$:
\begin{equation}\label{Diracequation}
   [\bm{\alpha}\cdot(\bm{p}-\bm{V})+\beta(m+S)
    +V-\bm{\omega}\cdot\hat{\bm{J}}]\psi_k=\epsilon_k\psi_k.
 \end{equation}
Here $\hat{\bm{J}}=\hat{\bm{L}}+\frac{1}{2}\hat{\bm{\Sigma}}$ is the total angular momentum of the nucleon spinors, and the fields $S$ and $V^\mu$ are connected in a self-consistent way to the densities and current distributions; for details, see Refs.~\cite{Peng2008Phys.Rev.C24313,Zhao2011Phys.Lett.B181}. The iterative solution of these equations yields single-particle energies, expectation values of the angular momentum, energy, quadrupole moments, etc. 

In this work, the energy density functional DD-ME2~\cite{Lalazissis2005Phys.Rev.C024312} is adopted. 
Since the level density of the single-particle levels for the present rod-shaped states is rather low, the cranking relativistic-Hartree-Bogoliubov calculations show that the pairing correlations could be neglected safely.
The calculations are free of additional parameters. Equation~(\ref{Diracequation}) is solved in a 3D Cartesian harmonic oscillator basis~\cite{Koepf1989Nucl.Phys.A61} with $N=12$ major shells to provide converged results. 


\begin{figure}[bp]
\centering
\includegraphics[width=6.5cm]{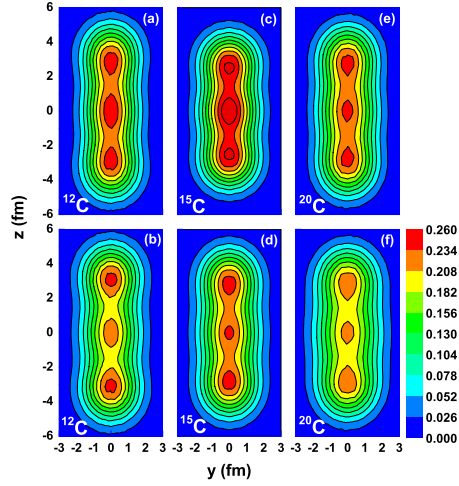}
\caption{(color online) Proton density distributions in the $y$-$z$ plane ($x$ direction is integrated) calculated by using the cranking covariant density functional theory for $^{12}$C, $^{15}$C, and $^{20}$C at the rotational frequencies $\hbar\omega=0.0$ MeV (a), (c), (e) and $\hbar\omega=3.0$ MeV (b), (d), (f).}
\label{fig2}
\end{figure}

\begin{figure*}[!htbp]
\centering
\includegraphics[width=17cm]{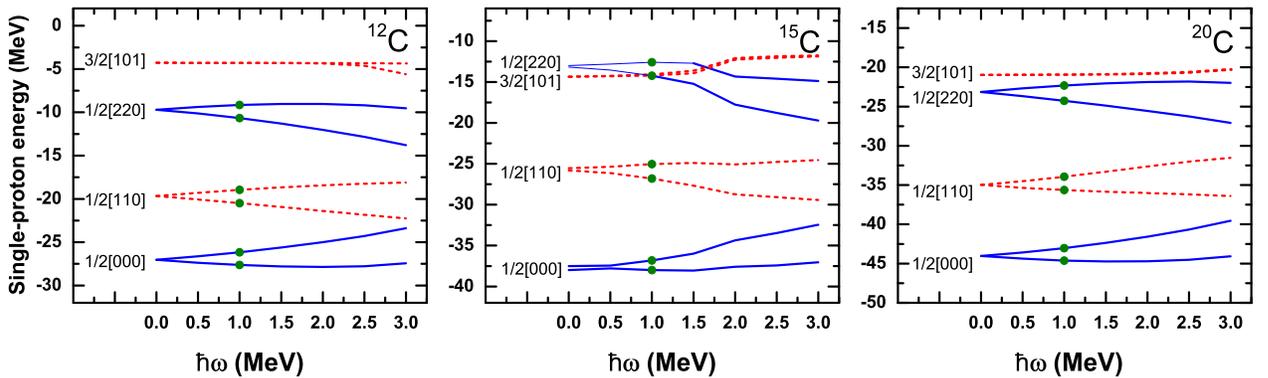}
\caption{(color online) Single-proton energies (in rotating frame) as functions of the rotational frequency for $^{12}$C, $^{15}$C, and $^{20}$C. Each orbital is labeled by the corresponding Nilsson quantum number of its maximal component. The solid and dashed lines denote the single-particle states with positive and negative parity, respectively. The solid circles denote the occupied orbitals. }
\label{fig3}
\end{figure*}

In the present calculations, we first solve Eq.~(\ref{Diracequation}) for $^{12}$C without rotation iteratively by assuming the initial fields $S$ and $V$ with a very large prolate deformation. In this way, one self-consistent solution with 3$\alpha$ linear-chain configuration for $^{12}$C has been obtained.
By taking the obtained potential as the initial potential, self-consistent calculations have been performed for C isotopes at various rotational frequencies. 
With the increase of spin and isospin, both the protons and neutrons are treated self-consistently by filling the orbitals according to their energy from the bottom of the well.
As a result, Fig.~\ref{fig1} shows the obtained expectation values of the angular momentum as functions of the rotational frequency for C isotopes from $A=12$ to $A=20$. 
One can easily classify these isotopes into three groups according to the behavior of their angular momenta. The first group contains $^{12,13,14}$C whose angular momenta are very close to each other.
It reveals from the linearly increasing tendency of the angular momenta that the moments of inertia are nearly constant; the slope is almost constant. The values of the moments of inertia are about 2.5 (MeV)$^{-1}\hbar^2$ which are very close to the corresponding classical values for a rigid rotor [around 3.0 (MeV)$^{-1} \hbar^2$].

The four nuclei $^{15,16,17,18}$C constitute the second group. Here, the backbending phenomenon, an abrupt increase of the moments of inertia, is shown clearly around $\hbar\omega=2.0$ MeV, which indicates some structure changes with the increasing angular momentum. It should be noted here that, for the lower spin part before the backbending, a rod-shaped solution requires an unchanged proton occupation which could be achieved by tracing the proton levels~\cite{Peng2008Phys.Rev.C24313,Zhao2012Phys.Rev.C54310} (see below).
By adding more neutrons, the third group is built with the nuclei $^{19}$C and $^{20}$C. Similar to the first group, the angular momenta here also increase linearly with the rotational frequency. 
This means that the moments of inertia here are nearly constant as well, but their values [around 4.6 (MeV)$^{-1}\hbar^2$] are much larger than that of $^{12,13,14}$C; this is due to the fact that the additional valence neutrons in $^{19}$C and $^{20}$C contribute more angular momentum to the system (see below).

Since the neutron number is changing for different C isotopes, it is convenient to show the structure of the rod-shaped C isotopes by using their proton density distribution. 
The proton density distributions for the same group differ only in barely visible detailed structures. Therefore, we show one sample for each group, i.e., $^{12}$C, $^{15}$C, and $^{20}$C, in Fig.~\ref{fig2}, illustrating the large deformation and the general structure produced by the three clusters. 
One could see that the extremely deformed rod-shape structure exists in all cases, and in particular an exotic structure of the 3$\alpha$-linear chain is very clearly seen.  

Because of the fact that the rod-shape structure in $^{12}$C persists with increasing spin and isospin as shown in Fig.~\ref{fig2}, it is important to check whether the proton configurations are stabilized against particle-hole deexcitations. 
To this end, the single-proton levels in the rotating frame together with their occupation are shown in Fig.~\ref{fig3} for the nuclei $^{12}$C, $^{15}$C, and $^{20}$C. 
Each level is labeled by the corresponding Nilsson quantum numbers $\Omega [N n_z \Lambda]$~\cite{Nilsson1955Mat.Fys.Medd.Dan.Vid.Selsk.1} of its maximal component, and positive and negative parities are represented by solid and dashed lines, respectively. It is worthwhile to mention that levels with small $\Lambda$ values correspond to densities close to the symmetry axis, while those with large $\Lambda$ values correspond to densities away from the symmetry axis.

For the nucleus $^{12}$C, all the levels are doubly degenerate at $\hbar\omega=0.0$ MeV and are split into two levels with increasing rotational frequency due to the violation of the time-reversal symmetry.
Moreover, the occupied proton levels here are always the lowest levels in energy from $\hbar\omega=0.0$ MeV to $\hbar\omega=3.0$ MeV. This indicates that the configuration is quite stable against any particle-hole deexcitations. 

For the case of $^{15}$C, however, the unoccupied proton level $3/2[101]$ gets lower than the occupied $1/2[220]$ level at a small rotational frequency, maybe due to the strong attractive interaction among protons and neutrons in the $p$ shell. 
 Note that here one has to trace the proton levels to stabilize the calculation, and thus the occupied $1/2[220]$ levels are indicated by thin lines in Fig.~\ref{fig3}.
This means that the linear configuration is not well stabilized, since the proton at the level $1/2[220]$ could easily jump to the level $3/2[101]$ to get lower energy. Nevertheless, the occupied level $1/2[220]$ is decreasing with the increasing frequency $\hbar\omega$ due to the Coriolis effect~\cite{Ring1980}.
When the frequency $\hbar\omega$ is larger than 2.0 MeV, this level becomes lower than the unoccupied level $3/2[101]$, and thus the configuration is getting stabilized. 

Similar to $^{12}$C, the configuration of $^{20}$C is also very well stabilized. The single-proton level scheme of $^{20}$C is very similar to that of $^{12}$C except for the magnitude of the energies. Because of the neutron-proton correlations, the single-proton energies of $^{20}$C are much lower than those of $^{12}$C. 

\begin{figure}[!htbp]
\centering
\includegraphics[width=8.5cm]{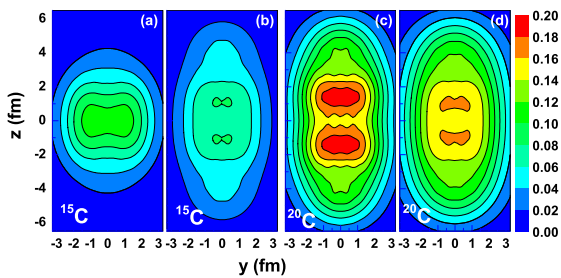}
\caption{(color online) Valence neutron distributions in the $y$-$z$ plane ($x$ direction is integrated) defined as the difference between the neutron and proton densities for $^{15}$C and $^{20}$C at the rotational frequencies $\hbar\omega=0.0$ MeV (a), (c) and $\hbar\omega=3.0$ MeV (b), (d).}
\label{fig4}
\end{figure}

The stability of the rod-shape states is strongly related to the valence neutrons, which are treated self-consistently by filling the neutron orbitals according to their energy.
In Fig.~\ref{fig4}, the valence neutron densities outside the core $^{12}$C, approximated as the difference between the neutron and proton densities $\rho_n-\rho_p$, for $^{15}$C and $^{20}$C are shown as examples. 

\begin{figure}[!htbp]
\centering
\includegraphics[width=7cm]{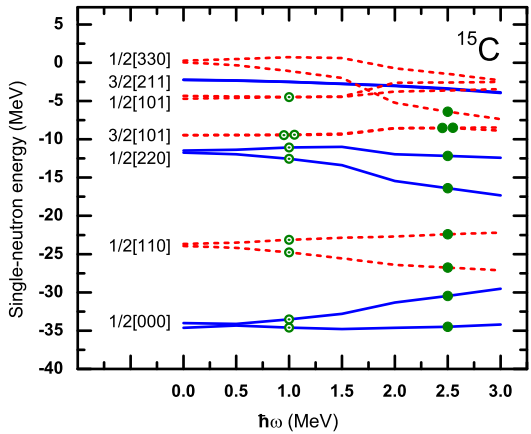}
\caption{(color online) Neutron single-particle energies (in rotating frame) as functions of the rotational frequency for $^{15}$C. The open and solid circles denote, respectively, the occupied orbitals before and after the level crossing near $\hbar\omega = 1.75$ MeV. }
\label{fig5}
\end{figure}

For $^{15}$C, the valence neutrons present an oblate distribution with two concentrations along the $y$ axis at $\hbar\omega=0.0$ MeV. Such a structure would hinder the formation of a rod shape along the $z$ axis, and thus it prevents the rod-shaped proton configuration from being stabilized. At $\hbar\omega=3.0$ MeV, however, the valence neutron changes to present a prolate distribution which is conducive to form a rod-shaped state, and thus the rod-shaped proton configuration could be well stabilized. Such a change essentially arises from the change of the occupation of the neutron orbitals as shown in Fig.~\ref{fig5}. Specifically, the $1/2[330]$ orbital drops rapidly with the rotational frequency and starts to be occupied at higher angular momentum. Such an orbital, usually called as a $\sigma$ orbital, would contribute a prolate distribution to the neutron density. 

The single-neutron levels of $^{20}$C have the same order in energy as that of $^{15}$C, and thus the $\sigma$ orbital $1/2[330]$ is always occupied even at $\hbar\omega=0.0$ MeV. This is quite helpful to the formation of the rod shape, and, as a result, the rod-shaped proton configuration of $^{20}$C could be well stabilized. 

The ground-state energies for the C isotopes from $A=12$ to $A=20$ have also been calculated in the present framework and compared with the data~\cite{Wang2012Chin.Phys.C1603}. It is found that the calculated ground-state energies are in very good agreement with the data with a root-mean-square deviation of around 2.7 MeV. 
One can also easily extract the excitation energies at $\hbar\omega=0.0$ MeV of the predicted rod-shaped states which are estimated to be in between around 13 and 18 MeV. These values are much lower than the values suggested in the previous work (around 25 MeV)~\cite{Itagaki2001Phys.Rev.C14301}. Note that the present microscopic calculations do not assume the existence of an alpha particle {\it a priori} and, thus, include more degrees of freedom in a larger model space.

Apart from the ground-state properties, it has been shown that, after the collective correlations are treated properly by angular momentum projection, the low energy spectroscopic properties of carbon isotopes can also be reproduced quite well~\cite{Yao2011Phys.Rev.C24306}. 
As a first-order approximation for the variation after angular momentum projection~\cite{Beck1970Z.Phys.26}, the cranking model has been extremely successful in the field of nuclear high spin phenomena for many years.
For cluster bands in light nuclei, as in Ref.~\cite{supplement}, very good agreement is achieved between the calculated two-alpha cluster bands in Be isotopes and the data.
Therefore, it would be very interesting to validate the present results obtained from the microscopic cranking CDFT in comparing them to experiment. 
The calculated energy spectra for C isotopes are given in Ref.~\cite{supplement} for a direct comparison with future experimental results.  Note that a moment of inertia of $\hbar^2/2 {\cal I}\sim$ 120 keV was reported by Freer {\it et al.} in Ref.~\cite{Freer2014Phys.Rev.C54324} for $^{14}$C. This moment of inertia just corresponds to our results when a neutron(s) occupies the $\sigma$ orbit around the 3$\alpha$ chain (green and red lines in Fig. 1 of Ref.~\cite{supplement}).
The 3$\alpha$ configurations slightly bent from the linear chain have been discussed in many works, e.g., in Refs.~\cite{Furutachi2011Phys.Rev.C21303,Baba2014Phys.Rev.C64319}, while not in the present work due to the fact that a cranking CDFT framework with octupole deformation is still not available up to date. 
However, at the bandhead, there are evidences that the rod shape in carbon isotopes could still be realized with the octupole degrees of freedom in both the nonrelativistic~\cite{Maruhn2010Nucl.Phys.A1} and relativistic~\cite{Ebran2014Phys.Rev.C54329} density functional theories.

In summary, we have discussed the rod-shaped configuration in C isotopes, which has been known to be very difficult to stabilize for a long time, by using the cranking covariant density functional theory.
The major advantages of the present framework include (i) the cluster structure is investigated without assuming the existence of clusters {\it a priori}, (ii) the nuclear currents are treated self-consistently, (iii) the density functional is universal for all nuclei throughout the periodic chart, and the present investigation is expected to be reliable and to have predictive power, and (iv) a microscopic picture can be provided in terms of intrinsic shapes and single-particle shells self-consistently.

Extreme isospin and spin are considered to be two key mechanisms for the stability of the rod-shaped configurations. In the present work, we have investigated the rod-shaped carbon isotopes, for the first time, by treating these two degrees of freedom simultaneously in a self-consistent and microscopic way. By increasing the isospin and/or spin, the appearance of the anomalously deformed rod shape can be clearly seen in the C isotopes. Through the effects from the Coriolis term, the $\sigma$ orbital, which is very important for the rod shape, comes down in energy and enhances the stability of the rod-shaped configuration with respect to the bending motion. Although this important neutron configuration for the rod shape ($\sigma$ orbitals) was the excited one in our early work~\cite{Itagaki2001Phys.Rev.C14301}, now it is shown to become the lowest one around the rod shape in the fast rotating frame.

\begin{acknowledgments}
This work is partly supported by the Chinese Major State 973 Program 2013CB834400, by the NSFC (Grants No. 11175002, No. 11105005, and No. 11335002), and by U.S. Department of Energy (DOE), Office of Science, Office of Nuclear Physics, under Contract No. DE-AC02-06CH11357. 
Numerical computation was carried out at the Yukawa Institute Computer Facility and the computing resources of the Laboratory Computing Resource Center at Argonne National Laboratory.
\end{acknowledgments}

%

\end{document}